\documentclass[aps,floatfix,showpacs,prl,twocolumn]{revtex4}
\usepackage{graphicx,bm}

\begin{document}
\title{Vortex crystallisation in classical field theory}
\author{Carlos Lobo, Alice Sinatra and Yvan Castin  \\ {\small Laboratoire
Kastler Brossel, Ecole Normale Sup\'{e}rieure, 24 Rue Lhomond, 75231
Paris Cedex 05, France}}

\begin{abstract}
We show that the formation of a vortex lattice in a weakly interacting
Bose condensed gas can be modeled with the nonlinear Schr\"{o}dinger
equation for both $T=0$ and finite temperatures without the need for an
explicit damping term. Applying a weak rotating anisotropic harmonic
potential we find numerically that the turbulent dynamics of the field
produces an effective dissipation of the vortex motion and leads to the
formation of a lattice. For $T=0$ this turbulent dynamics is triggered by
an already known rotational dynamic instability of the condensate. For
finite temperatures, noise is present at the start of the simulation and
allows the formation of a vortex lattice at a lower rotation frequency,
the Landau frequency. These two regimes have different vortex
dynamics. We show that the multimode interpretation of the classical field
is essential.
\end{abstract}

\maketitle

\begin{figure}[t]
\begin{tabular}{ccc}
{} & \hspace{0.3cm} $\Omega=0.8\omega$ \ \ \  $T=0$  \hspace{1cm}
$\Omega=0.6\omega$ \ \ \ $T=8 \hbar\omega$ & {} \\
\rotatebox{90}{\small \phantom{.} \hspace{1cm} $t=4000\omega^{-1}$ 
 \hspace{2cm} $t=740\omega^{-1}$ \hspace{2cm}
$t=453.3\omega^{-1}$ \hspace{2.5cm} $t=0$} &
\includegraphics[width=8.0cm,height=16cm,clip=]{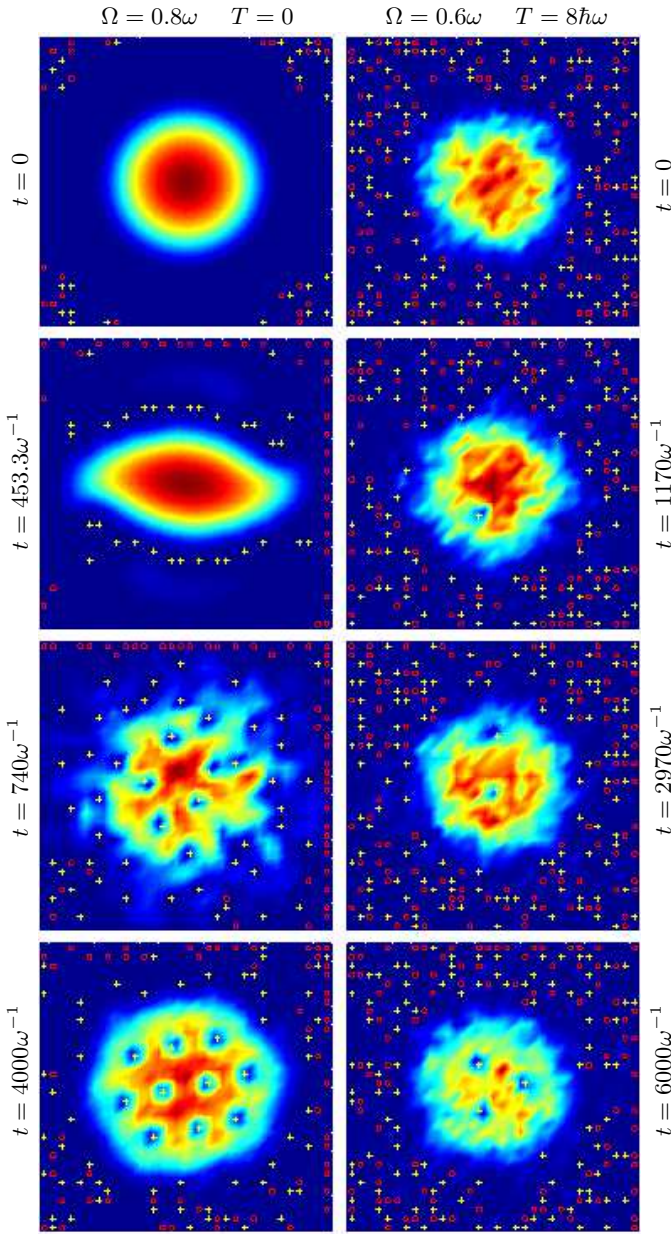} &
\rotatebox{90}{\small \phantom{.} \hspace{1cm} $t=6000\omega^{-1}$ 
\ \ \ \hspace{1.7cm} $t=2970\omega^{-1}$ \ \ \ \hspace{1.5cm}
$t=1170\omega^{-1}$ \hspace{2.5cm} $t=0$}
\end{tabular}
\caption{Cut along the radial plane ($z=0$) of the system spatial
density at different
times. Crosses (circles) indicate position of vortices of
positive (negative) charge
\cite{note3}.  Left column: $T=0$, $\Omega_f=0.8\omega$.  Top to bottom:
initial state; near instability; turbulent
behaviour; end of simulation. Right column: $k_B T=8\hbar \omega$,
$\Omega_f=0.6\omega$. Top to bottom: initial state; entry of first vortex; 
entry of second vortex;
end of simulation with a 3-vortex lattice.}
\label{fig:T08}
\end{figure}

\begin{figure}
\centerline{\includegraphics[width=7cm,height=4.5cm,clip=]{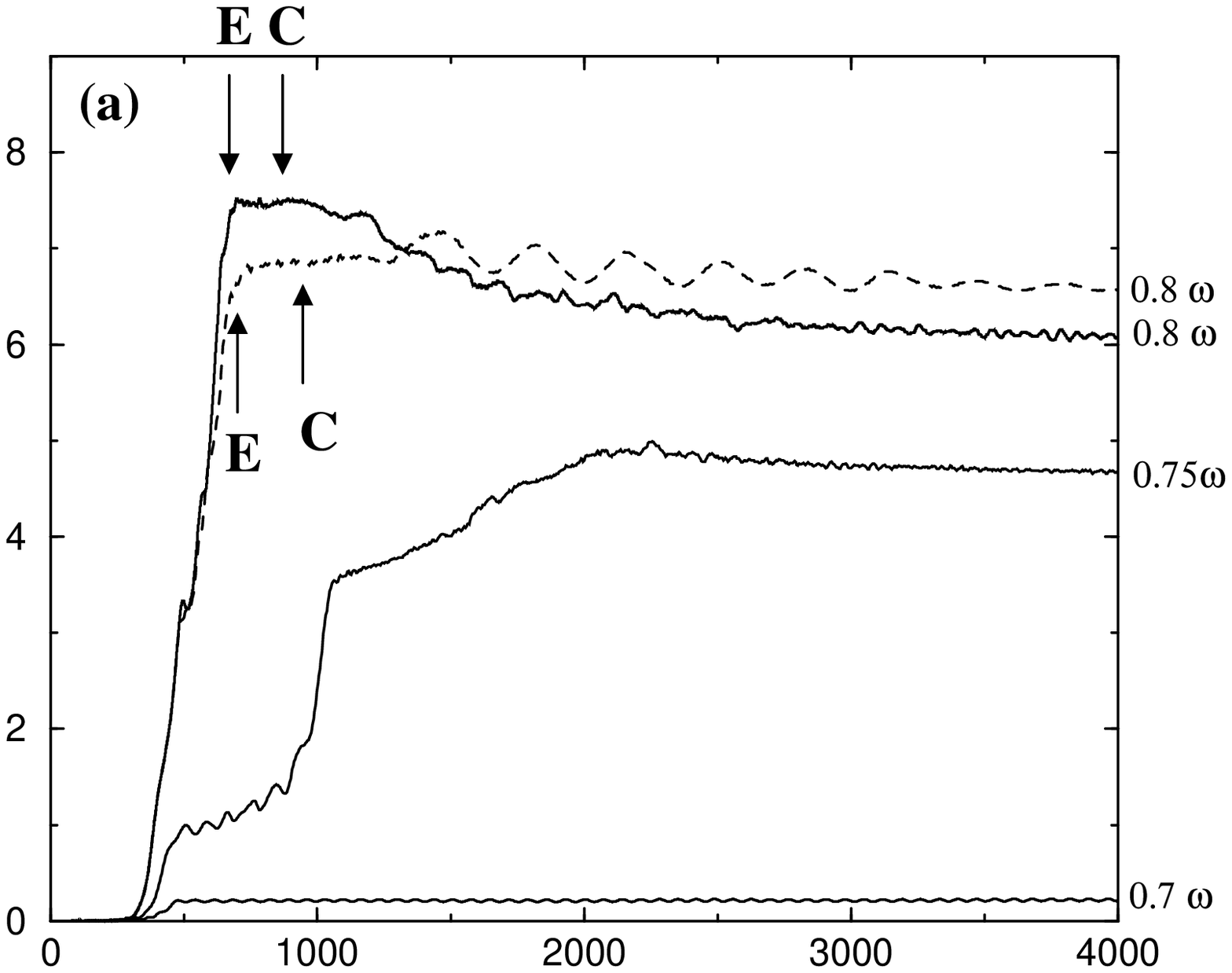}}
\centerline{\includegraphics[width=7cm,height=4.5cm,clip=]{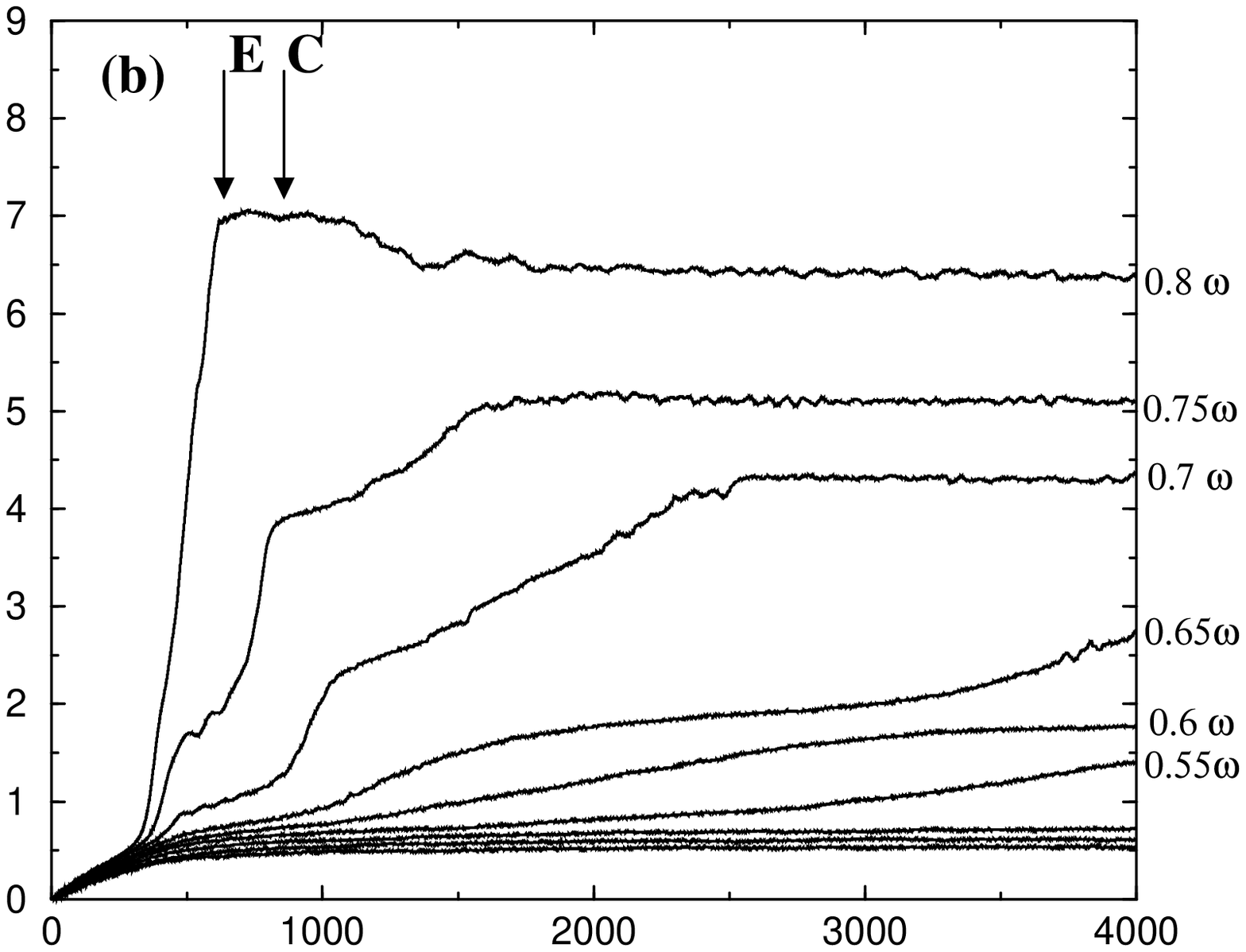}}
\centerline{\includegraphics[width=7cm,height=4.5cm,clip=]{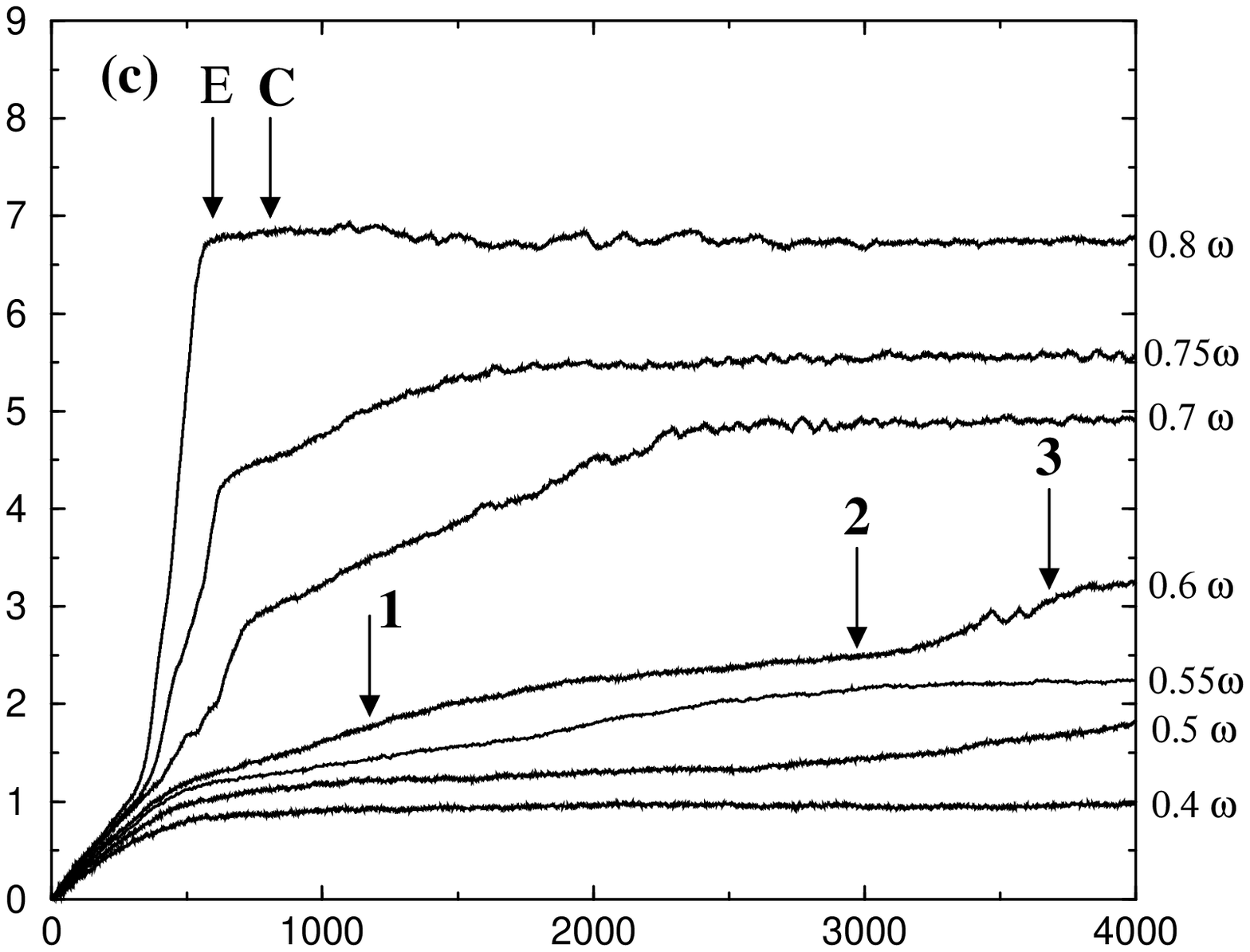}}
\caption{Angular momentum in units of $\hbar$ per atom. The arrows marked
E and C indicate the entry of the
vortices into the condensate and the crystallisation of the lattice 
for $\Omega_f=0.8\omega$.  (a) $T=0$, solid
lines from bottom to top:  $\Omega_f/\omega=0.7(0),0.75(7),0.8(10)$;
dashed line: $\Omega_f/\omega=0.8(10)$ with a gridsize of $64 \times 64
\times 256$. All other curves were done on a $32 \times 32 \times 128$
grid. In parenthesis is the number of vortices in the lattice at the end
of the simulation. (b) $k_B T=4 \hbar \omega$,
$\Omega_f/\omega=0.4(0),0.45(0),0.5(0),0.55(1),
0.6(1),0.65(2),0.7(6),0.75(7),\\0.8(10)$. (c) $k_B T=8 \hbar \omega$,
$\Omega_f/\omega=$0.4(0), 0.5(1), 0.55(1), 0.6(3), 0.7(7), 0.75(7), 0.8(10). The
arrows correspond to the approximate entry time of the vortices for
$\Omega_f=0.6 \omega$ as shown in Fig.\ref{fig:T08}. Note that the angular
momentum shows no signature of the entries.}
\label{fig:lz}
\end{figure}
Several groups have recently observed the formation of a vortex lattice in
weakly interacting Bose gases \cite{ENS,MIT,JILA,Foot}. Theoretically
there have been attempts to understand the formation process
\cite{Feder,Ueda,Gardiner,Lundh} with simulations of the Gross-Pitaevskii
equation for the condensate wavefunction. These papers either do not
consider the effect of the noncondensed modes \cite{Feder,Lundh} or they
approximate it with an added damping term. All of them stress the need for
explicitly including this term since the vortices have to lose energy to
be able to relax to a lattice configuration.

Here we consider this problem within the framework of the classical theory
of a complex field whose {\it exact} equation of motion is the standard
nonlinear Schr\"{o}dinger equation (NLSE). Within this framework we can
give the true numerical answer which can be used as a testing ground for
approximations such as the single mode one of \cite{Ueda,Gardiner} to
study the formation of the vortex lattice. If the dynamics of the field
were ergodic we would expect that, in a fast enough rotating external
potential, the vortex lattice would be formed without the need for any
dissipative term. The issue of relaxation to thermal equilibrium for
purely Hamiltonian dynamics is analogous to the Fermi-Ulam-Pasta problem
\cite{Fermi} and has been studied also in the context of Bose gases
\cite{Svistunov,Sachdev,Burnett,Castin}. Here we study the formation of
the lattice in three dimensions from an initially nonrotating Bose
condensed gas both at $T=0$ and at finite temperatures using the NLSE. We
find that, contrary to the prevalent view, the lattice is formed in both
cases without any need for a damping term which suggests that
thermalisation takes place in our system.

We start our simulations with the nonrotating classical field in thermal
equilibrium. For $T=0$ the initial state is simply a pure condensate, that
is, with a field proportional to the condensate wavefunction $\phi$ given
by the Gross-Pitaevskii equation in the absence of rotation, $\psi =
\sqrt{N_0}\phi$ where $N_0$ is the condensate atom number. For finite
temperatures we sample the initial thermal equilibrium in the Bogoliubov
approximation at a given temperature $T$ for a fixed number $N_0$ of
condensate particles.  In this approximation the classical field is given
by $\psi({\bf r},0) = \sqrt{N_0}\phi({\bf r})+\psi_\perp({\bf r})$. The
random field $\psi_\perp({\bf r})$ orthogonal to $\phi$ \cite{Cartago}
representing the thermal noise is given by
\begin{equation} 
\psi_\perp({\bf r})=\sum_n b_n u_n({\bf r})+b_n^* v_n^*({\bf r}) 
\end{equation}
where the $u_n$ and $v_n$ are the Bogoliubov mode functions associated
with $\phi$ and the $b_n$ are independent random c-numbers taken from a
Gaussian distribution that obeys the classical equipartition formula,
 $\langle b_n^* b_n
\rangle=k_BT/\epsilon_n$, $\epsilon_n$ being the Bogoliubov energy of mode
$n$.  In practice, to sample this distribution we use the Brownian motion
method described in \cite{Cartago}. In our work, the field $\psi$ is to be
interpreted not as the condensate wavefunction but as the whole matter
field (unlike \cite{Feder,Gardiner,Ueda,Lundh}). We present here results
from single realisations of the field $\psi$ which
experimentally correspond to single runs. We have checked that
different realisations lead to similar results.

In our simulations we consider a Bose condensed gas initially trapped in a
cigar-shaped harmonic potential with oscillation frequencies whose ratio
is 1:1:0.25, with $10^5$ atoms of mass $m$ and a coupling constant
$g=0.0343$ in units of $\hbar \omega a_0^3$ where $\omega$ is the radial
frequency and $a_0=\sqrt{\hbar/m \omega}$ is the oscillator length. The
corresponding chemical potential is $\mu=8 \hbar \omega$.  We start each
simulation with the gas in thermal equilibrium. We abruptly turn on the
trap anisotropy which leads to a change in the radial frequencies:
$\omega^2_{x,y}=\omega^2(1\mp \epsilon)$ where $\epsilon=0.025$.  Then the
rotation frequency $\Omega(t)$ of this anisotropy is slowly increased from
zero to a final value $\Omega_f$ over $500 \omega^{-1}$, to follow
Procedure I in \cite{Subhasis}. After that we let the gas evolve in the
presence of the rotating anisotropy until, for most of our simulations,
the angular momentum of the gas reached a steady state.

The calculation
is performed in the rotating frame so that
the NLSE takes the form
\begin{equation}
i\hbar\partial_t\psi = 
\left[-\frac{\hbar^2}{2m}\Delta + U(\mathbf{r}) + g |\psi|^2
-\Omega(t) L_z
\right]\psi
\end{equation}
where $L_z$ is the angular momentum operator along $z$ and $U$ is the anisotropic
harmonic potential. The field $\psi$ is subject to
periodic boundary conditions in the rotating frame.
Our gridsize is $32\times32\times128$ corresponding to an energy cutoff of
$32 \hbar \omega$ per spatial direction, although we have also run
simulations on a $64\times64\times256$ grid (see below). 
We have checked that it is necessary to have time independent boundary conditions
in the rotating frame: periodic boundary conditions in the lab frame
arrest the rotation of the non-condensed gas. 
We have also checked that a pure condensate cannot be set into rotation
by the effect of the boundary conditions only,
since the condensate density is extremely
weak at the borders of the grid for our choice of gridsize. 
The harmonic trap anisotropy is then a crucial element for the formation
of the lattice at $T=0$ by triggering the dynamic instability of 
\cite{Subhasis}.

{\it Zero initial temperature:} This set of simulations can be divided
into two groups: those for which the final rotation frequency is
$\Omega_f/\omega \leq 0.7$ and those with $\Omega_f/\omega \geq 0.75$.
Between these two values lies the threshold for the dynamic instability of
the condensate which changes the subsequent dynamics dramatically
\cite{Subhasis}. In the first group, as the rotation frequency gradually
increases with time, the condensate adiabatically follows a steady state,
apart from excitations of the surface modes leading to a very small
periodic variation of the angular momentum (see curve for
$\Omega_f=0.7\omega$ in Fig.\ref{fig:lz}a). With increasing $\Omega_f$,
the condensate's final state becomes more and more elliptically deformed,
surrounded by a ring of vortices which however never enter it. The second
group shows completely different behaviour when $\Omega(t)/\omega \simeq
0.75$ (see left column of Fig.\ref{fig:T08}): the instability sets in, the
condensate becomes slightly S-shaped at $t\simeq 450\omega^{-1}$ before
being highly deformed and undergoing very turbulent motion \cite{Feder}.  
This is accompanied by a large increase in angular momentum of the gas
from almost zero when $\Omega(t)<0.75\omega$ to between $5 \hbar$-$7\hbar$
per particle (see Fig.\ref{fig:lz}a). At this point ($t\simeq 670
\omega^{-1}$) several vortices enter the high density region and, in less
than 200$\omega^{-1}$, settle down to form a well-defined lattice. After
this, a period of relaxation of around 800$\omega^{-1}$ begins with the
initially rotating lattice finally stopping in the rotating frame. There
remains a small random motion of the vortices around their
equilibrium positions in
the lattice together with density fluctuations in and around the
condensate.

At the end of the simulation, damping of the vortex motion has occurred
and the initial energy of the vortex motion has been transferred in an
effectively irreversible way to other degrees of freedom of the field.  A
similar phenomenon has been observed for the relative motion of two
condensates \cite{Castin}. {\it If we assume that the field has reached a
thermal distribution}, we can calculate the temperature of the system by
taking the final state of the simulation and evolving it with the
conjugate gradient method in a trap rotating at $\Omega_f$. This reduces
its energy and takes it to the local minimum associated with the vortex
lattice. We then calculate the energy difference $\Delta E$ between the
final state of the simulation and the one at the minimum. Assuming that
Bogoliubov theory is valid, $\Delta E$ must correspond to the energy of a
classical thermal distribution of weakly coupled harmonic oscillators of
amplitude $b_n$ which obeys the equipartition formula $\langle b_n^*b_n
\rangle\epsilon_n=k_BT$, with $n$ being the Bogoliubov mode number.  So,
if ${\mathcal N}$ is the number of modes in the system (and keeping in
mind that we have to subtract the one corresponding to the condensate) 
then we have
\begin{equation}
\Delta E=\sum_n \langle b_n^*b_n \rangle\epsilon_n=({\mathcal N}-1)k_BT.
\label{eq:classical}
\end{equation}
The final temperature is
$0.616 \hbar \omega$ for $\Omega_f=0.75\omega$ and $0.754 \hbar
\omega$ for $\Omega_f=0.8\omega$, in other words it is extremely
small, less than a tenth of the chemical potential.

We have also carried out a simulation on a larger grid
($64\times64\times256$) to check the dependence on size. We chose
$\Omega_f=0.8\omega$ and compared it with the one on the
$32\times32\times128$ grid. The vortex nucleation and crystallisation
phases are very similar and occur at roughly the same times. At longer
times two differences arise: first, there are large underdamped
oscillations of the angular momentum (see Fig.\ref{fig:lz}a). An analysis
of the simulation suggests that these oscillations are those of the
scissors mode.  Second, the final temperature ($0.094 \hbar \omega$)
differs by the ratio of the number of modes as expected:
at time $t=500\omega^{-1}$ when $\Omega(t)=\Omega_f$, $\psi$
had not yet reached the boundary in the smaller grid case and so the
evolution of $\psi$ on both grids was identical up to this time with the
same total energy which was conserved at later times resulting in the same
value of $\Delta E$. This exemplifies the fact that, in classical field
theories, the relationship between energy and temperature depends on the
energy cutoff.

Since the thermal occupation of the modes is directly proportional to the
temperature, we expect that all relaxation processes which involve
scattering from or into those modes (such as Landau-Beliaev damping) will
be reduced. We are thus led to the conclusion that, for our simulations
starting at $T=0$, relaxation rates in the equilibration period after the
formation of the lattice could depend on the size of the grid. However,
with the present numerical results, we were not able to demonstrate this.

{\it Finite initial temperature:} We performed simulations starting with
$k_B T=4\hbar \omega$ and $k_B T=8 \hbar \omega$.  Now not only the
condensate but also other modes are occupied in the initial state, with a
thermal distribution. For a final rotation frequency below that of the
dynamic instability, the situation is quite different from that of the
zero temperature case: the condensate is never deformed and the vortices
do enter the condensate if 
$\Omega_f \geq 0.55 \omega$ for $k_B T=4 \hbar \omega$ 
and if $\Omega_f \geq 0.5 \omega$ for $k_B T=8 \hbar \omega$. 
We have checked numerically that $0.51 \omega$ is the Landau critical 
frequency above which the vortex-free condensate is no longer a
local minimum
of the energy. During the real time
evolution of the right column of Fig.\ref{fig:T08} corresponding to
$\Omega_f=0.6\omega$, we find that the vortices enter only one at a time.
That is, as the angular momentum of the cloud increases, one vortex, out
of the group of vortices that surrounds the condensate will enter it and
spiral slowly clockwise towards the center on a time scale of hundreds of
$\omega^{-1}$. After that vortex has reached the center, a second one
enters slowly, repeating the trajectory of the first until it starts to
interact with it and the two orbit around each other for a while after
which a third will enter.  At the end of the simulation, coinciding with
the achievement of the plateau in angular momentum, the lattice becomes
stationary in the rotating frame and no further vortex enters the
condensate (see right column of Fig.\ref{fig:T08}). For
$\Omega_f=0.7\omega$ we find that the condensate deforms itself
elliptically after which three vortices enter at the same time and form a
rotating lattice. After that, and spaced by several hundred $\omega^{-1}$,
a fourth and then a fifth vortex enter. Finally, two further vortices
enter simultaneously to form the final seven vortex lattice. At each
intermediate stage there is always a well defined lattice present although
it is not stationary in the rotating frame.  We should contrast this with
the scenario of \cite{Ueda,Gardiner} where a large number of vortices
enter all at once into the condensate in a ring configuration and then
some of them form a lattice while others are shed and leave the
condensate.

For $\Omega_f$ above the dynamic instability frequency the situation is
quite similar to the corresponding one at $T=0$. Once the instability has set
in the lattice is formed for both temperatures in about $200\omega^{-1}$ as
in the $T=0$ case (see Fig.\ref{fig:lz}b,c-this weak temperature
dependence was also found experimentally by \cite{Raman}).
The main difference is that the relaxation time
for the lattice to stop rotating is much shorter, on the order of a
hundred $\omega^{-1}$, not eight hundred.

It is important to emphasize the multimode interpretation of the field.
Transposing Penrose and Onsager's definition to the classical field
theory, the condensate wavefunction is defined as the eigenvector
corresponding to the largest eigenvalue of the one-body density matrix
$\langle \psi^*({\bf r}') \psi({\bf r}) \rangle$ where the average is over
an ensemble of initial states. If the system becomes turbulent because it
encounters an instability, the trajectories of the neighboring
realisations will diverge exponentially. However, after averaging, we
believe that the condensate wavefunction will not be a turbulent
function. For $T=0$ there is only one initial state and so we replace
ensemble averaging by one over time in the steady state regime \cite{LL}.
In our simulations with $\Omega_f=0.8\omega$ the system must therefore be
understood as becoming intrinsically multimode even though we started at
$T=0$ with a pure condensate.

{\it Conclusions:} We have identified two very different regimes for the
crystallisation of the vortex lattice in the classical field theory. At
$T=0$ when the dynamic instability sets in (for high
enough $\Omega_f$), the vortices enter the condensate abruptly and settle
into a lattice even if there is no dissipative term in the equation of
motion. This scenario is consistent with the one found experimentally at
the ENS \cite{ENS2} with comparable timescales. At finite temperatures,
where \cite{Ueda,Gardiner} add an explicit damping term, we see that the
thermal classical field can provide the damping of the vortex motion on
its own which leads to very different behaviour from that observed by
those authors in that the vortices can enter one by one into the
condensate and settle into a lattice before the entry of the following
one. So far there has been no experimental check of this scenario. 
Finally, as the $T=0$ case shows, any theoretical model which
singles out the condensate mode for separate treatment with a
Gross-Pitaevskii-type equation could run into trouble in turbulent
situations since the separation between condensed and non-condensed modes
would be hard to keep.

We have been informed that crystallisation
of the vortex lattice has also been observed in a simulation without a
damping term by the group of N. Bigelow. We thank B. Durin, L. Carr, I.
Carusotto, G. Shlyapnikov and J. Dalibard for useful contributions. C.
L. acknowledges a fellowship from the Funda\c{c}\~ao para a Ci\^encia e
Tecnologia of Portugal. LKB is a unit of ENS and of Universit\'e Paris 6
associated to CNRS.


\begin{thebibliography}{99}
\bibitem{ENS} K. W. Madison, F. Chevy, W. Wohlleben, and J. Dalibard,
Phys. Rev. Lett. 84, 806 (2000).
\bibitem{MIT} J. R. Abo-Shaeer, C. Raman, J. M. Vogels, and W. Ketterle,
Science {\bf 292}, 476 (2001).
\bibitem{JILA} P. C. Haljan, I. Coddington, P. Engels, and E. A. Cornell,
Phys. Rev. Lett. 87, 210403 (2001).
\bibitem{Foot} E. Hodby, G. Hechenblaikner, S. A. Hopkins, O. M.
Marag\'{o}, C. J. Foot, Phys. Rev. Lett. {\bf 88}, 010405 (2002).
\bibitem{Feder} 
D. L. Feder, A. A. Svidzinsky, A. L. Fetter, C. W. Clark,
Phys. Rev. Lett. {\bf 86}, 564 (2001).
\bibitem{Ueda} M. Tsubota, K. Kasamatsu, M. Ueda, Phys. Rev.
A {\bf 65}, 023603 (2002);
K. Kasamatsu, M. Tsubota, M. Ueda, cond-mat/0211394.
\bibitem{Gardiner}  A. A. Penckwitt, R. J. Ballagh, C. W. Gardiner,
cond-mat/0205037.
\bibitem{Lundh}  E. Lundh, J.-P. Martikainen, K.-A. Suominen,
cond-mat/0211401.
\bibitem{Fermi} E. Fermi, J. Pasta, and S. Ulam, {\it Collected Papers of
Enrico Fermi} (edited by E. Segr\'e, University of Chicago Press, Chicago,
1965).
\bibitem{Svistunov} B. V. Svistunov, J. Moscow. Phys. Soc. {\bf 1}, 373
(1991).
\bibitem{Sachdev} K. Damle, S. M. Majumdar, and S. Sachdev, Phys. Rev. A
{\bf 59}, 5037 (1996).
\bibitem{Burnett} R. J. Marshall, G. New, S. Choi, and K. Burnett, Phys.
Rev. A {\bf 59}, 2085 (1999).
\bibitem{Castin} A. Sinatra, P. Fedichev, Y. Castin, J. Dalibard, and G.
V. Shlyapnikov, Phys. Rev. Lett. {\bf 82} 251 (1998).
\bibitem{Cartago}  A. Sinatra, C. Lobo, Y. Castin, J. Phys. B
{\bf 35}, 3599 (2002).
\bibitem{Subhasis} S. Sinha, Y. Castin, Phys. Rev. Lett.
{\bf 87}, 190402 (2001).
\bibitem{P}  Yu. Kagan and B. Svistunov, Phys. Rev. Lett. {\bf
79} 3331 (1997). B. M. Caradoc-Davies, R. J.  Ballagh, and K. Burnett, Phys. Rev. Lett.
{\bf 83}, 895 (1999).
\bibitem{note3} Vortex positions are found by integrating the gradient of
the phase around each grid square in the plane. Vortices in the $T=0$
initial state are due to incomplete convergence to the groundstate.  
\bibitem{Raman} J.R. Abo-Shaeer, C. Raman, W. Ketterle, Phys. Rev. Lett.
{\bf 88}, 070409 (2002).
\bibitem{LL} L. D. Landau and E. M. Lifshitz, {\it Statistical Physics}
3rd edition, Part 1 \S 1, Pergamon Press (Oxford, 1980).
\bibitem{ENS2} K. Madison, F. Chevy, V. Bretin, and J. Dalibard, Phys.
Rev. Lett. {\bf 86}, 4443 (2001).
\end{thebibliography}
\end{document}